\documentclass{elsart3}

\usepackage{times}

\usepackage{graphicx}

\usepackage{amssymb}
\begin{document}

\begin{frontmatter}


\title{Critical Exponents from General Distributions of Zeroes}

\author[ia1]{Wolfhard Janke,}
\author[ia2]{Des Johnston,}
and
\author[ia3]{Ralph Kenna\corauthref{cor3}}
\address[ia1]{Institut f\"ur Theoretische Physik, Universit\"at Leipzig, 
                Augustusplatz 10/11, 04109 Leipzig, Germany}
\address[ia2]{Department of Mathematics, Heriot-Watt University, Riccarton, 
               Edinburgh, EH14~4AS, Scotland}
\address[ia3]{School of Mathematical and Information Sciences, Coventry University, 
               Coventry, CV1~5FB, England}
\corauth[cor3]{Corresponding Author:}
\ead{r.kenna@coventry.ac.uk}

\begin{abstract}
All of the thermodynamic information on a statistical mechanical system is encoded in the 
locus and density of its partition function zeroes.
Recently, a new technique was developed which enables the extraction of the latter  
using finite-size data 
of the type typically garnered from a computational approach.
Here that method is extended to deal with more general cases. 
Other  critical points of a type which 
appear in many models are also studied.
\end{abstract}

\end{frontmatter}

\section{Introduction}
\label{sec1} 
%
Phase transitions are of central interest in statistical physics and related fields. 
Second-order transitions are signaled by divergences and characterised 
by critical exponents (e.g., $\alpha$ for the specific heat and $\nu$ 
for the correlation length in the temperature driven case). 
Such non-analytic behaviour is only present in systems of infinite 
extent and is therefore inaccessible to Monte Carlo 
simulations, which are restricted to a finite number of degrees of freedom.

Traditionally, however, finite-size scaling (FSS) may 
be used to extract thermodynamic information from such systems. 
FSS, based on the hypothesis that there are only two relevant length scales
(namely the correlation length of the infinite 
system and its finite-size counterpart), 
typically only allows determination of {\em{ratios}} 
of critical exponents associated with 
thermodynamic functions, 
such as $\alpha/\nu$. 
An exception is 
the correlation length critical exponent $\nu$ 
which 
can be directly extracted 
from logarithmic derivatives of magnetization moments or from the slope of 
the Binder parameter $\sim \langle m^{2k} \rangle / \langle m^k \rangle^2$, 
$k=1, 2, \dots$. Alternatively one may also consider the scaling 
behaviour of pseudocritical points. The latter, defined 
as the extrema of 
thermodynamic functions, approach the transition point as $L^{-\lambda}$, where 
$L$ denotes the linear extent of the system and $\lambda$ is the so-called shift 
exponent. The exponent $\lambda$ coincides with $1/\nu$ in many models, but this 
is not a consequence of FSS and is not always true. See, e.g., \cite{JaKe02} for 
a review of the recent literature concerning this point. 
A further complication that arises from the latter approach is that such a 
fit involves three parameters and is non-linear, so usually is quite unstable 
and often inaccurate.

An increasingly popular approach is the use of FSS of the zeroes of the partition function.
FSS of the lowest zeroes in the complex temperature plane (Fisher zeroes)
provides a direct and accurate method to extract the exponent $\nu$, 
and the imaginary parts of the lowest zeroes (labelled by an index $j$)
scale with lattice extent as ${\rm{Im}}z_j \sim L^{-1/\nu}$.
The real part of the lowest partition function zero is another pseudocritical point,
generally scaling as $L^{-\lambda}$.

It has long been known that a full understanding of the properties of the bulk system 
requires knowledge of the {\em{density}} of zeroes too. 
Until recently, determination of the density from finite-size Monte Carlo data was considered 
difficult if not impossible.
The source of the difficulties is that it involves reconstruction 
of a continuous density function from a discrete data set as the 
density of zeroes for a finite system essentially consists of a 
set of delta functions. 

Recent considerations have bypassed these difficulties by 
focusing instead on the integrated density of zeroes \cite{UsJSP}. 
In particular, this new approach facilitates measurement of the 
strength of the transition through {\em{direct}} determination of $\alpha$
(as opposed to traditional FSS measurements of the ratio $\alpha / \nu$). 
While the new technique proved successful, it was limited to systems where the zeroes fall on curves 
in the complex parameter plane and where the zeroes are non-degenerate.
While these two properties are common to most models in statistical physics,
they are not generic and a host of examples now exist where the zeroes
are distributed across a two-dimensional region and/or occur in degenerate sets. 
Here, the new technique is extended to deal with such general distributions of zeroes \cite{UsNPB}.

\section{General Distributions of Zeroes}
\label{sec2} 
%
When the partition function, $Z_V$, for a  system of volume $V=L^d$
($d$ being the dimensionality of the system)
can be written as a polynomial in an appropriate function, $z$, of temperature, field or 
of a coupling parameter, one has 
$
Z_V(z)
\propto
\prod_j{(z-z_j(V))}
$
where $j$ labels the zeroes. 
In the general case where the distribution of zeroes is two-dimensional, 
the free energy may be expressed as
$
f_V
=
\int\!\! \int
g_V (x,y)
\ln{(z-z_c-x-iy)}
dxdy
$,
where $g_V$ is the density of zeroes and 
$(x,y)$ give their location in the complex plane with the 
critical point, $z_c$, as the origin.
In the infinite-volume case, Stephenson has shown that 
the density near a second-order transition point satisfies a certain homogeneous partial
differential equation, the solution of which may be written as
$
 g_\infty (x,y) = y^{1-\alpha - m} f(x/y^m)
$,
where   $m$ is related to the shape of the 
locus \cite{Stephenson}. Integrating out the $x-$direction, and 
integrating up to a point $r$ in the $y-$direction  gives the cumulative density there to be
\begin{equation}
 G_\infty (r) \propto r^{2-\alpha}
\,.
\label{get}
\end{equation}
From this expression, the exponent $\alpha$ may be directly measured provided that
a sensible definition for the cumulative density of zeroes can be applied to a finite system.
Such a function is defined as follows. If the $j^{\rm{th}}$ zero
is $n$-fold degenerate the densities to its immediate left and right
are given by
$V G_V(r) = j-1$ and $j+n-1 $ respectively.
The density {\em{at}} the $j^{\rm{th}}$  zero, $r_j$, is then defined as an average:

\begin{equation}
 G_L(r_j) = \frac{1}{V} \left(j+\frac{n}{2}-1 \right)   
\label{soln}
\, .           
\end{equation}
Combined with (\ref{get}), this allows direct determination of the critical exponent $\alpha$.

\section{Applications}
\label{sec3} 
We apply the new technique to Ising models in two dimensions for which the zeroes are 
calculable and which possess each of the new features we wish to encapsulate.
In each case, the real, physical, critical point is characterized by
$\alpha = 0$.
%
%

{\bf{Brascamp-Kunz lattice with anisotropic couplings:}}
The finite-size, standard, nearest-neighbour, square lattice Ising model has been solved in two dimensions 
for certain sets of boundary conditions including those first studied by 
Brascamp and Kunz \cite{BK}. There, periodic boundary conditions are used
in one direction, while at the extremities in the other direction the spins are fixed to
$+$ on the one hand and the alternating sequence $+-+-\dots$ on the other.
\begin{figure}[t]
\begin{center}
\includegraphics[scale=0.25]{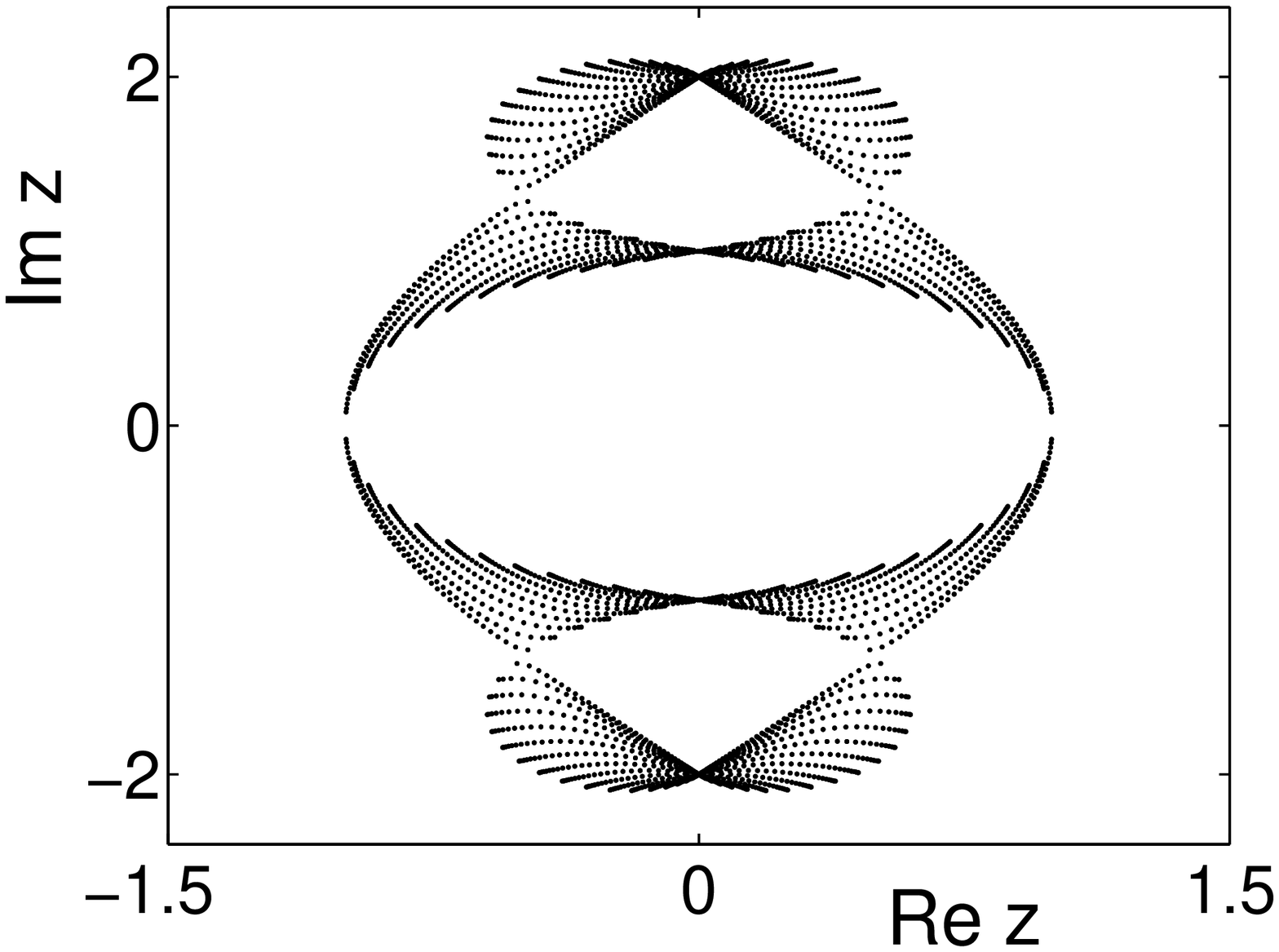} 
\includegraphics[scale=0.25]{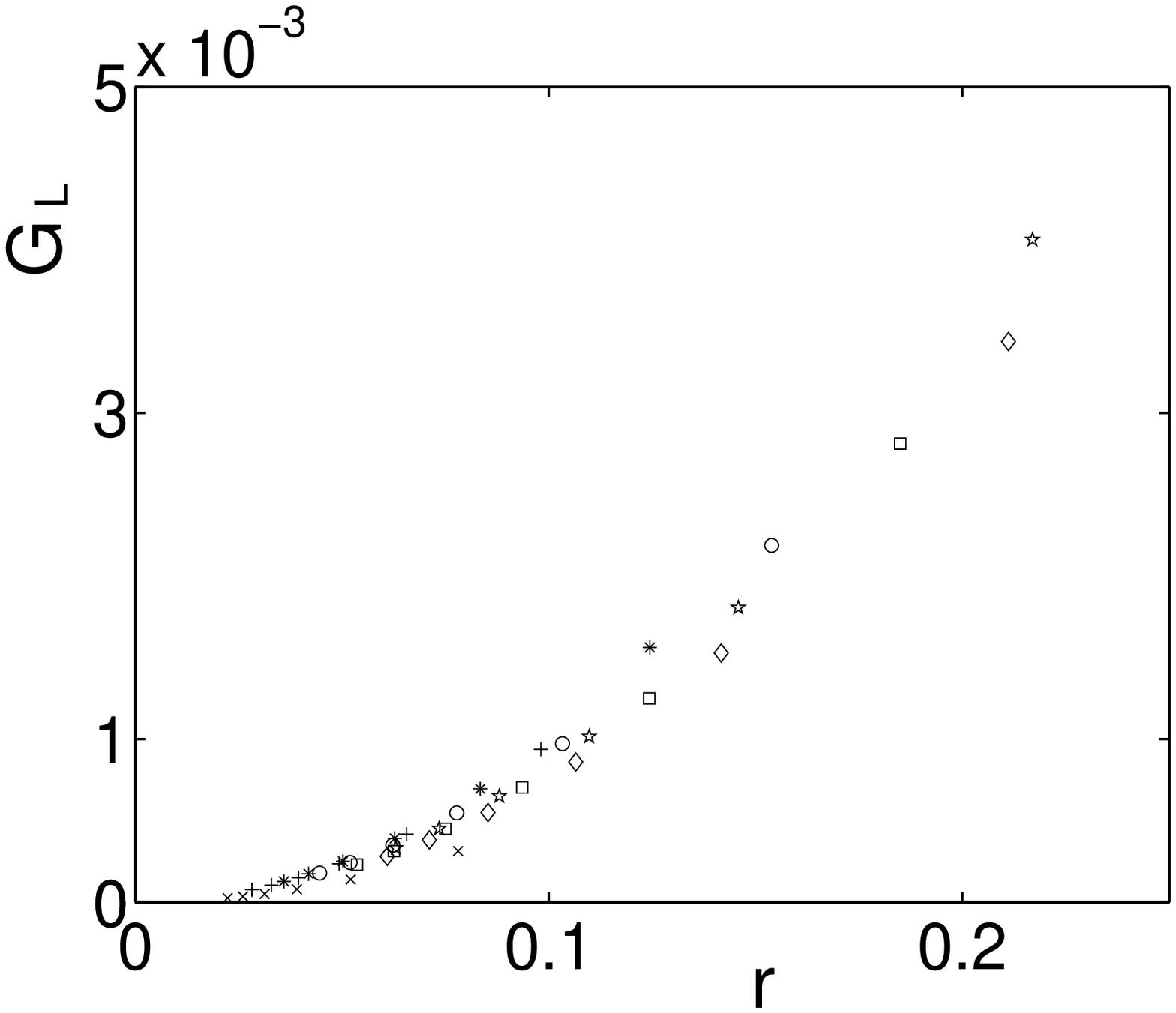} 
\caption{The zeroes (top) for the anisotropic $L=40$ Brascamp-Kunz Ising model 
with $J^\prime/J=3$ and
their density distribution (bottom) near $z=1$ for $L=40$--$140$ 
and $j=1$ ($\times$), 
     $j=2$ ($+$),
     $j=3$ ($*$),
     $j=4$ ($\circ$),
    $j=5$ ($\Box$),
    $j=6$ ($\diamond$),
     $j=7$ ({\tiny{$\bullet$}}) and
     $j=8$ ($\star$).
 } 
\end{center}
\end{figure}
For a lattice of linear extent $L$
with anisotropic couplings $J$ and $J^\prime$,
the partition function then takes the form of a single product 
(as opposed to a sum of four such products, which is the case when periodic boundary conditions 
in both directions are used \cite{Ka49}),
greatly ameliorating the computation of its zeroes \cite{JaKe02}. 
The zeroes are easily determined numerically and
are distributed across a two-dimensional region  in 
the $z=2\sinh( 2 \beta)$ plane as shown in 
Fig.~1 for $J^\prime/J=3$ (with $\beta = J/k_B T$).     
The zeroes impact onto the real axis at the point $z=1$ and the
critical behaviour is  dominated by the
zeroes close by. 
The cumulative density distribution for this set of zeroes
is also plotted in Fig.~1.
That the curve goes through the origin indicates the 
presence of a transition and an appropriate fit  yields 
$\alpha = -0.016(32)$ compatible with expectations.
%
%

{\bf{Bathroom-tile lattice:}}
Two-dimensional distribution of zeroes may also be obtained from systems with isotropic couplings
as demonstrated  in \cite{MaSh95} where the system is described in detail.
In principle the full finite-size partition function is a sum of four
product terms. 
One may construct Brascamp-Kunz type boundary conditions for this lattice, which
would have the effect of projecting out one of these terms in the expression
for the partition function. Alternatively, and more conveniently, one may
discretize one of the terms in the partition function with periodic
boundary conditions and assume for the purposes of the analysis herein 
that the scaling behaviour of that term is generic. Indeed, since 
we are essentially interested in testing the scaling of the cumulative density of zeroes
rather than formulating the finite lattice models themselves, this is
sufficient for our purposes (see also \cite{UsNPB}).

\begin{figure}[t]
\begin{center}
\includegraphics[scale=0.25]{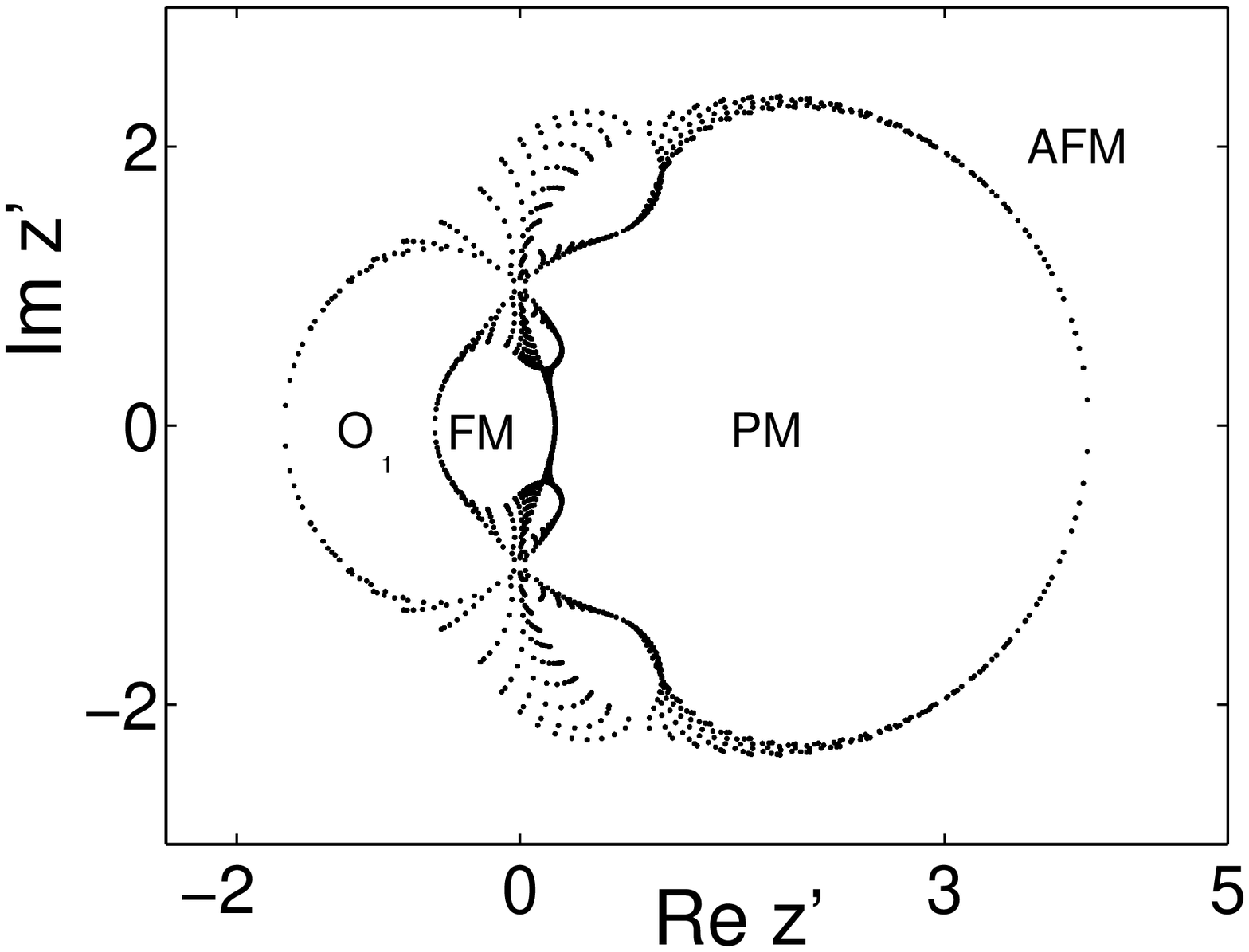} 
\includegraphics[scale=0.25]{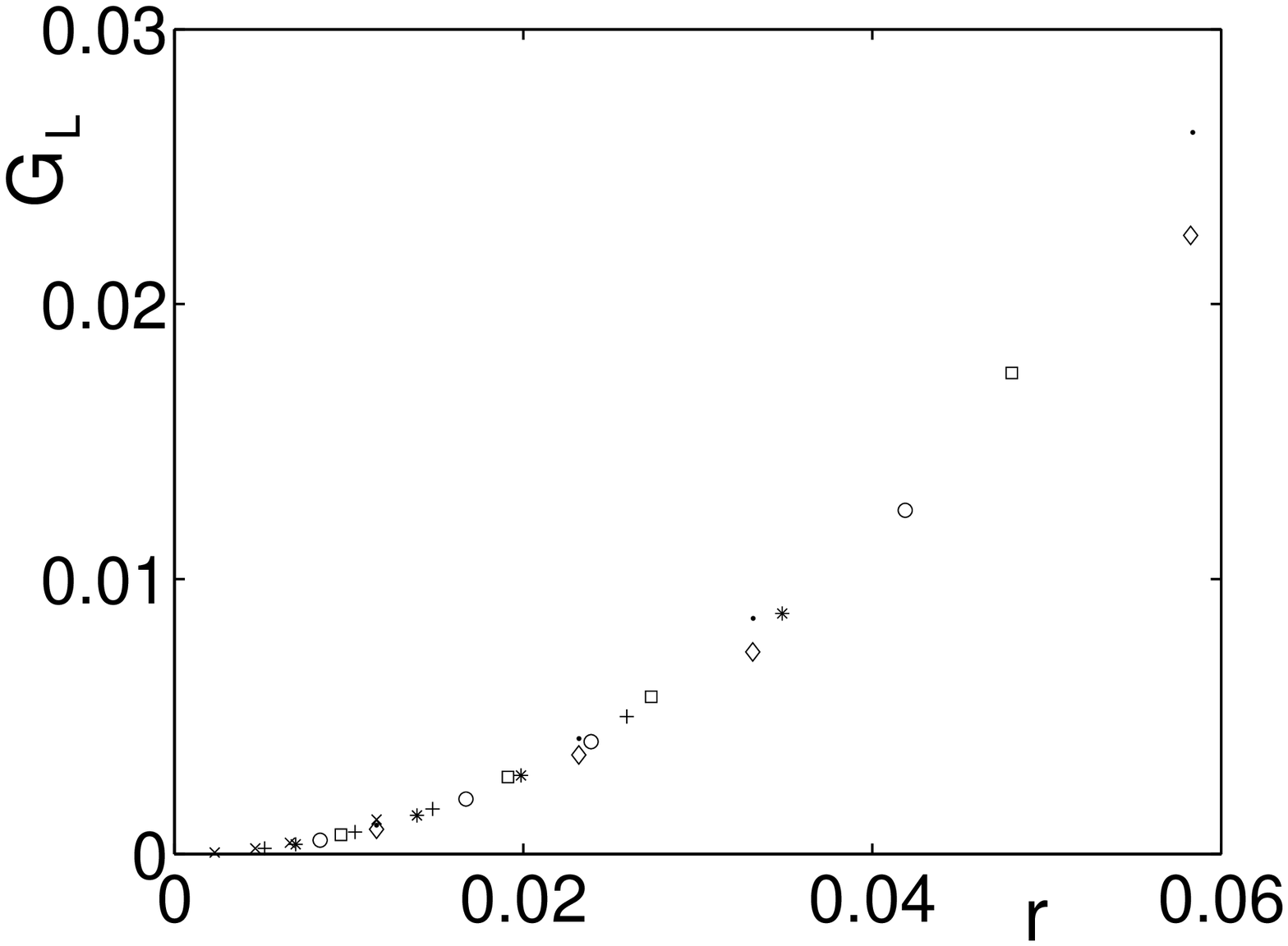} 
\caption{The  zeroes (top) 
for the $L=40$ bathroom-tile Ising
model and 
their density (bottom) near the ferromagnetic critical point
from data with $L=40$--$200$ 
and  $j=1$-$4$ ($\times$), 
     $j=5$-$12$ ($+$),
     $j=13$-$16$ ($*$),
     $j=17$-$24$ ($\circ$),
    $j=25$-$32$ ($\Box$),
     $j=33$-$40$ ($\diamond$) and
     $j=41$-$44$ ({\tiny{$\bullet$}}).
} 
\end{center}
\end{figure}
The zeroes  of such a term  have varying degrees of degeneracy and are 
depicted in the complex $z^\prime=\exp{(-2 \beta)}$ plane
in Fig.~2,
where AFM, PM, FM and O$_1$ indicate the 
anti-ferromagnetic,
paramagnetic, ferromagnetic and unphysical phases, respectively.
The physical ferromagnetic critical point is given by 
$z^\prime =  0.249\,038\,4\dots$ and the cumulative density of zeroes nearby
is also depicted in the figure.
A fit yields $\alpha = 0.002(18)$,
consistent with zero.
There is also an antiferromagnetic transition point at
$z^\prime = 4.015\,445\,4\dots$. A density fit to the zeroes nearby 
yields $\alpha = 0.0006(163)$, again compatible with $\alpha=0$.

Since the finite-size  partition function is known exactly in this case, 
and is a convenient single product, it is
possible to analytically extract the $\nu$ and $\lambda$ exponents 
from conventional FSS of the lowest lying zeroes. Indeed, one finds that
$\nu=1$ (which again gives $\alpha=0$ through hyperscaling) 
and $\lambda = 2$ in both the ferromagnetic and antiferromagnetic cases.

%
%

{\bf{Complex vertices:}}

It has been pointed out that unphysical singular
points (i.e., points for which there is no real $\beta$)
may be considered as ordinary critical points 
with distinctive critical exponents (see, e.g., \cite{MaSh95}
and references therein).

\begin{figure}[t]
\begin{center}
\includegraphics[scale=0.25]{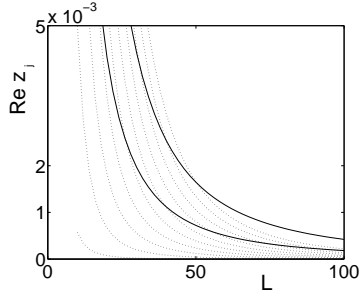} 
\caption{Finite-size dependency of the zeroes close to 
the complex singular point $z=i$ in the isotropic Brascamp-Kunz model.
The dotted lines scale as 
${\rm{Re}} z_j \sim L^{-3}$ and the solid lines as ${\rm{Re}} z_j \sim L^{-2}$.
} 
\end{center}
\end{figure}
For the example of the two-dimensional Ising
model with Brascamp-Kunz boundary conditions,
the complex vertex points at $z=i$ and $2i$ in Fig.~1 are thus also of interest.
However, analysis shows these not to be of the conventional type. 
Fig.~3 is a  plot showing how  the first few zeroes in the 
simpler isotropic model approach such a vertex at $z=i$ with increasing lattice size.

To understand the unusual behaviour depicted, an analytic approach is required.
A full scaling analysis of this model at its real transition point $z=1$
is given in \cite{JaKe02}. For a square lattice of linear extent $L$,
the zeroes, which are labelled $z_{ij}$, are given by 
\begin{eqnarray}
{\rm{Re}}z_{ij} & = & \frac{1}{2}
\left( \cos{\theta_i} + \cos{\phi_j}
\right)
\,,
\nonumber
\\
{\rm{Im}}z_{ij} & = & \sqrt{1- ({\rm{Re}}z_{ij})^2 }
\,,
\label{funny}
\end{eqnarray}
where $\theta_i = (2i-1)\pi/L$ and $\phi_j=j\pi/(L+1)$.
Expanding the cosines for large $L$ gives ${\rm{Re}}z_{ij} 
                                = 1 + 
                               {{\mathcal{O}}}(L^{-2})$
and ${\rm{Im}}z_{ij} = {\mathcal{O}}(L^{-1})$, recovering
$\lambda=2$ and $\nu=1$, as above.
For ${\rm{Re}}z_{ij} $ close to $1$ (the physical critical point), both
$i$ and $j$ have to be close to zero and the above expansion is legitimate.
However, close to the unphysical point $z=i$, the two cosines in 
(\ref{funny}) have to cancel and the expansion is no longer valid.
Cancellations of this type 
remove the leading $L^{-2}$ term for the approach to the vertex 
 of the first few zeroes 
(which then 
scale as $L^{-3}$)
and do not occur at a real critical point. 
This explains the odd scaling behaviour in Fig.~3
and  demonstrates the dangers inherent to a restricted 
traditional analysis of leading zeroes.
This and related issues will be elaborated upon elsewhere.

The situation  close to the unphysical  points 
$z^\prime = - 0.601\,231\,8\dots$ and
$z^\prime =-1.663\,251\,9\dots$ in Fig.~2 
is more conventional and density analyses
yield $\alpha = 0.007(12)$ and $\alpha = -0.0095(123)$ respectively.
The exponents $\nu$ and $\lambda$ may also be extracted 
analytically here and one again finds $\nu=1$, $\lambda=2$. 
It is interesting to note that these are yet more  cases where 
$\nu$ does not coincide with $\lambda$  \cite{JaKe02}.

\bibliography{density2D}
\end{document}